\documentstyle[12pt]{article}
\title{Chiral condensate of lattice QCD with massless quarks from the probability distribution function method}
\vspace{2cm}
\author{Xiang-Qian Luo\\
{\small\sl CCAST (World Laboratory), P.O. Box 8730, 
Beijing 100080, China}\\
{\small\sl Department of Physics, Zhongshan (Sun Yat-Sen) University, 
Guangzhou 510275, China\thanks{Mailing address. Email: stslxq@zsu.edu.cn}}\\
}

\date{\today}

\begin{document}
\maketitle

\begin{abstract}
We apply the probability distribution function method 
to the study of chiral properties of  QCD with quarks
in the exact massless limit. A relation among the chiral condensate, zeros of the Bessel function and eigenvalue of Dirac operator is also given.
The chiral condensate in this limit can be measured 
with small number of eigenvalues of the massless Dirac operator and without any ambiguous mass extrapolation. 
Results for SU(3) gauge theory with quenched  Kogut-Susskind  quarks on the $10^4$ lattice are shown.
\end{abstract}

\vspace{2cm}

\centerline{PACS number(s): 11.30.Qc, 11.15Ha, 11.30.Rd}

\vspace{2cm}

\centerline{Published in {\it Physical Review} {\bf D69}, 076012 (2004).}

\newpage

\section{INTRODUCTION}

One can not completely understand physics of hadrons 
without understanding the QCD vacuum, 
which main properties 
are confinement and spontaneous chiral-symmetry breaking, characterized by the 
non-vanishing chiral condensate in the massless (chiral) limit. 
Suppose the quarks are degenerate. The chiral condensate per  flavor is 
\begin{eqnarray}
\langle {\bar \psi} \psi \rangle 
={1 \over N_cV} \langle {\rm Tr} \Delta^{-1} \rangle ,
\label{thermo}
\end{eqnarray}
where $N_c=3$ is the number of colors, $V$ is the number of
lattice sites and $\Delta$ is the fermionic matrix. The trace is taken in the color, spin and position space.

On a finite lattice,  however,
direct computation of the chiral condensate from a chiral symmetric action
leads to $\langle {\bar \psi} \psi \rangle \vert_{m=0}=0$, even though there would be spontaneous chiral-symmetry breaking
in the infinite volume limit.
In standard lattice simulations,
one has to add a mass term $\sum_x m {\bar \psi} \psi$ to the action and calculate 
$\langle {\bar \psi} \psi \rangle$ at some set of non-zero bare fermion mass $m$, 
and then extrapolate $\langle {\bar \psi} \psi \rangle$
to the massless limit 
by means of some modeled fitting functions
(e.g., linear function, polynomial or logarithmic corrections).
Unfortunately, such a process might not be well justified, and sometimes it gives very different results. Here are two well known evidences.

\noindent
(a) QED in 4 dimensions. Non-compact lattice QED 
experiences a second order chiral phase transition at finite
some bare coupling constant $g$, where the chiral condensate vanishes.
The critical coupling and critical index determined by a $m$ extrapolation of 
$\langle {\bar \psi} \psi \rangle$ are ambiguous.
Detailed discussions can be found in Ref.\cite{QED4}. 

\noindent
(b) The one-flavor massless Schwinger model. 
In the continuum, as is analytically known,
$\langle {\bar \psi} \psi \rangle_{cont}
\approx 0.16 e$, where $e$ is the electric charge.
Unfortunately, a more careful study\cite{QED2} indicates the value of the
chiral condensate computed by mass extrapolation depends strongly on the mass range to extrapolate,
and the fitting function mentioned above.

In Ref.\cite{PDF}, an alternative, named the probability distribution 
function (p.d.f.) method was proposed to investigate
spontaneous chiral-symmetry breaking\footnote{One can find 
an analog in statistical physics, e.g. in  the analysis of spin glasses\cite{Sibani,Mezard}. However, to our knowledge, it is the first time
to apply such an idea to first principle quantum field theory, i.e., lattice gauge theory with fermions.}.
This method has been tested in the Schwinger model\cite{PDF},
and applied to the study of the spontaneous P and CT symmetry breaking\cite{Azcoiti:1999rq} and theta-vacuum like systems\cite{Azcoiti:2002vk}
as well as the phase transition of SU(2) lattice gauge theory at finite density\cite{Aloisio:2000rb}.

In this paper, we will further explore the p.d.f. method by 
applying it to SU(3) lattice gauge theory with  Kogut-Susskind (KS) fermions.
The rest of the paper is organized as follows. In Sec. \ref{sec_PDF}, we elaborate the idea of the p.d.f. method
and derive some relations between the eigenvalues of the Dirac operator and chiral condensate.
In Sec. \ref{sec_Data}, we present the results from numerical simulations.

\section{P.D.F. OF THE CHIRAL CONDENSATE}
\label{sec_PDF}

Let us characterize each vacuum state by $\alpha$ and 
the chiral condensate  by $\langle {\bar \psi} \psi \rangle_{\alpha} $.
The p.d.f. of the chiral condensate in the Gibbs state is defined by 
\begin{eqnarray}
P(c)=\sum_{\alpha} w_{\alpha} \delta \left(c- \langle {\bar \psi} \psi \rangle_{\alpha} \right),
\label{definition}
\end{eqnarray}
with $w_{\alpha}$  the weight to get the vacuum state $\alpha$. 
$P(c)$ is the probability to get the value $c$ 
for the chiral condensate from a randomly chosen vacuum state. 
If there is exact chiral symmetry in 
the ground state, $P(c)=\delta(c)$.
If chiral symmetry is spontaneously broken,
$P(c)$ will be a more complex function. Therefore, 
from the shape of the function $P(c)$ computed 
in the configurations generated by a chiral {\it symmetric} action with exact $m=0$, 
one can qualitatively judge whether chiral symmetry is spontaneously
broken.

In quantum field theory with fermions, 
chiral-symmetry breaking is dominated 
by the properties of the fermion fields 
under global chiral transformation $\psi \to \exp(i \mbox{\boldmath $\alpha \cdot \tau$}\gamma_5) \psi$,
with $\mbox{\boldmath $\tau$}$ the generator of the chiral symmetry group. 
From Eq. (\ref{definition}),
one can  define the p.d.f. of the chiral condensate
for a single gauge configuration $U$:
\begin{eqnarray}
P_U(c) ={ \int \left[ d{\bar \psi} \right] \left[ d \psi \right] 
~  \exp \left(-S_f \right) 
 \delta \left(c- {1 \over N_cV}\sum_{x} {\bar \psi}(x) \psi(x)\right) \over
\int \left[ d{\bar \psi} \right] \left[ d \psi \right] 
~  \exp \left(-S_f \right)} ,
\label{D1}
\end{eqnarray}
where in ``$\sum_{x} {\bar \psi}(x) \psi(x)$", 
summation over color,  spin and position indices is implied.

One can also define the p.d.f. for all gauge configurations
\begin{eqnarray}
P(c) = 
\bigg \langle \delta \left(c- 
{1 \over N_cV}\sum_{x} {\bar \psi}(x) \psi(x)\right)
\bigg \rangle ,
\label{pdf}
\end{eqnarray}
where the expectation value ``$< ...>$" is computed with the integration 
measure associated with the partition function
\begin{eqnarray}
Z = \int \left[ d{\bar \psi} \right] \left[ d \psi \right] \left[ dU \right]
~  \exp \left(-S \right) 
=\int \left[ dU \right] ~ \exp \left(-S_g + \ln {\rm det} \Delta \right) ,
\label{partition}
\end{eqnarray}
with $\Delta$ the fermionic matrix.

To study the chiral properties quantitatively,
it is more convenient to employ the Fourier transformed p.d.f.
\begin{eqnarray}
{\tilde P}_U(q) &=& \int^{\infty}_{-\infty}
dc ~ \exp \left(-iqc \right) P_U(c),
\label{fourier00}
\end{eqnarray}
and
\begin{eqnarray}
{\tilde P}(q) &=& \int^{\infty}_{-\infty}
dc ~ \exp \left(-iqc \right) P(c).
\label{fourier0}
\end{eqnarray}

Inserting Eq. ({\ref{pdf}) into Eq. (\ref{fourier0}),
one obtains
\begin{eqnarray}
{\tilde P}(q) &=& \int^{\infty}_{-\infty}
dc ~ \exp \left(-iqc\right) P(c)
\nonumber \\
&=& {1 \over  Z} \int \left[ d{\bar \psi} \right] \left[ d\psi \right] \left[ dU \right] 
\exp \left(-S_g-S_f\right) \int^{\infty}_{-\infty}
dc ~ \exp \left(-iqc\right) 
\delta \left(c- {1 \over N_cV} \sum_x  {\bar \psi} (x) \psi (x)\right)
\nonumber \\
&=& {1 \over  Z} \int \left[ d{\bar \psi} \right] \left[ d\psi \right] \left[ dU \right] 
\exp \left(-S_g+\sum_{x,y} {\bar \psi} (x) \left(\Delta_{x,y} - i {q \over N_cV}\delta_{x,y} \right) \psi (y) \right). 
\label{A1}
\end{eqnarray}
Integrating out the fermion fields, Eq. (\ref{A1}) becomes
\begin{eqnarray}
\nonumber \\
{\tilde P}(q) &=&  {1 \over Z} \int \left[ dU \right] 
{\rm det} \left(\Delta -{iq \over N_c V} I \right) \exp \left(-S_g \right)
\nonumber \\
&=&  {1 \over Z} \int \left[ dU \right] 
{{\rm det} \left(\Delta -{iq \over N_c V} I \right) \over   {\rm det} \Delta}
\exp \left(-S_g + \ln {\rm det} \Delta\right),
\label{A2}
\end{eqnarray}
where $I$ is the identity matrix. Generally, a fermionic matrix $\Delta$ can be decomposed as
\begin{eqnarray}
\Delta=m I+i \Gamma .
\end{eqnarray}
Denoting $\lambda_j$ the $j-$th positive eigenvalue of $\Gamma$, then the determinants in Eq. (\ref{A2}) are
\begin{eqnarray}
{\rm det} \Delta &=& \prod_{j=1}^{N_cV/2} \left(\lambda_j^2+m^2 \right) ,
\nonumber \\
{\rm det} \left(\Delta -{iq \over N_c V} I \right)
&=&\prod_{j=1}^{N_cV/2} \left(\lambda_j^2+m^2- {q^2+2imqN_c  V 
\over \left(N_cV \right)^2 } \right) .
\end{eqnarray}
Substituting them into Eq. (\ref{A2}), one obtains
%
\begin{eqnarray}
{\tilde P}(q)&=&  
{1 \over Z} \int \left[ dU \right] ~ \exp \left(-S_g + \ln {\rm det} \Delta \right)
\prod_{j=1}^{N_cV/2} \left( 1- {q^2+2imqN_c  V 
\over \left(N_cV\right)^2 \left(\lambda_j^2+m^2\right) } \right)
\nonumber \\
&=&  
\bigg \langle 
\prod_{j=1}^{N_cV/2} \left( 1- {q^2+2imqN_c  V 
\over \left(N_cV\right)^2 \left(\lambda_j^2+m^2\right) } \right)
\bigg \rangle .
\label{fourier}
\end{eqnarray}

From Eq.(\ref{fourier}), we derive relations Eq. (\ref{ps1}) and Eq. (\ref{ps3}) between the chiral condensate
and the eigenmodes of the Dirac operator.

\subsection{First relation}

According to the definition Eq. (\ref{fourier0}), the Fourier transformed p.d.f. of the chiral condensate can also
be written as
\begin{eqnarray}
{\tilde P}(q) &=& \int^{\infty}_{-\infty}
dc ~ \exp \left(-iqc \right) P(c)
\nonumber \\
&=&  
\int^{\infty}_{-\infty}
dc ~ \exp \left(-iqc \right) \bigg \langle\delta \left(c -{1 \over N_c V} \sum_x {\bar \psi}\psi  \right) \bigg \rangle
\nonumber \\
&=&  
\bigg \langle \exp \left(-{i q \over N_c V}\sum_x {\bar \psi}\psi  \right) \bigg \rangle .
\end{eqnarray}
Its derivative with respective to $q$ at $q=0$ is
\begin{eqnarray}
{\partial {\tilde P}(q) \over \partial q} \bigg \vert_{q=0} 
= -i\bigg \langle {1 \over N_cV} \sum_x {\bar \psi}\psi \bigg \rangle .
\label{B1}
\end{eqnarray}
On the other hand, using Eq. (\ref{fourier}), the derivative of ${\tilde P}(q)$ 
with respective to $q$ at $q=0$ is
\begin{eqnarray}
{\partial {\tilde P}(q) \over \partial q} \bigg \vert_{q=0} 
=
- i\bigg \langle {1 \over N_cV}\sum_{j=1}^{NcV/2} {2m 
\over \lambda_j^2+m^2}
\bigg \rangle .
\label{B2}
\end{eqnarray}
Comparing Eq. (\ref{B1}) and Eq. (\ref{B2}), 
we have
\begin{eqnarray}
\langle {\bar \psi} \psi \rangle 
=  
{1 \over N_cV} \bigg \langle \sum_{j=1}^{NcV/2} 
{2m \over \lambda_j^2 +m^2} \bigg \rangle . 
\label{ps1}
\end{eqnarray}
In the derivation, we haven't used the specific properties of lattice fermion formulation.
It is just the standard one in the literature. 
The disadvantage is that to get the chiral condensate in the chiral limit $\lim_{m \to 0} \lim_{V \to \infty} 
\langle {\bar \psi} \psi \rangle$, 
it requires a $m$ extrapolation, since $<{\bar \psi}\psi>\vert_{m=0}=0$ on a finite lattice.
As mentioned above, the result depends on the choice of the fitting function,
in particular near the chiral phase transition.
It also requires the calculation of all eigenvalues of the Dirac operator.
When the lattice volume is
large, the computational task is huge and not so
feasible. From Eq. (\ref{ps1}), one can also derive the so called Banks-Casher formula
\begin{eqnarray}
\lim_{m \to 0} \lim_{V \to \infty} 
\langle {\bar \psi} \psi \rangle 
= \lim_{\lambda \to 0} 
\pi \langle \rho(\lambda) \rangle ,
\label{ps2}
\end{eqnarray}
which is also frequently used in the literature. Here $\rho(\lambda)$ is the eigenvalue density.
This formula means that the eigenvalues relevant for chiral-symmetry breaking go 
to zero in the infinite volume limit.
The disadvantage is that on a finite lattice,  $\rho(0)=0$,  it requires a $\lambda \to 0$ 
extrapolation using some modeled fitting function.  Only  in the infinite volume limit,
the number of the eigenvalues approaching zero
diverges, so that $\rho (0) \neq 0$.

\subsection{Second relation}


In a theory with continuous U(1) chiral symmetry (exactly when $m=0$), 
the vacuum is characterized by an angle $\alpha \in [-\pi,\pi]$ and
the chiral condensate can be parameterized as
$\langle {\bar \psi} \psi \rangle_{\alpha}=c_0 \cos \alpha$, where
$c_0$ is the amplitude 
of the chiral condensate corresponding to the spontaneously broken continuous U(1) symmetry. 
According to the definition Eq. (\ref{definition}),  the p.d.f. of the chiral condensate is
\begin{eqnarray}
P(c)\bigg \vert_{m=0}
&=&\sum_{\alpha} w_{\alpha} \delta \left(c- \langle {\bar \psi} \psi \rangle_{\alpha} \right)
={1 \over 2 \pi} \int_{-\pi}^{\pi} d \alpha ~ \delta \left(c-c_0 \cos \alpha \right)
\nonumber\\
&=&
\bigg \{
\begin{array}{c}
{1 \over \pi \sqrt{c_0^2 -c^2}},   ~~~~~~~ c\in \left[-c_0,c_0\right],\\ 
0,  ~~~~~~~~~~~~~~~~~  c<-c_0, ~ {\rm or} ~ c>c_0. 
\end{array}
\label{C1}
\end{eqnarray}
The Fourier transformed p.d.f. is then
\begin{eqnarray}
{\tilde P}(q) \bigg \vert_{m=0}
&=& \int^{\infty}_{-\infty}
dc ~ \exp \left(-iqc\right) P(c) 
\nonumber \\
&=&  
{1 \over 2 \pi} \int_{-\pi}^{\pi} d \alpha ~ \int^{\infty}_{-\infty} dc  ~\delta \left(c-c_0 \cos \alpha \right) \exp \left(-iqc\right) 
\nonumber \\
&=&  
{1 \over 2 \pi} \int_{-\pi}^{\pi}  d\alpha ~ \exp \left(-iqc_0 \cos \alpha \right)=J_0 \left(qc_0\right),
\label{C2}
\end{eqnarray}
where $J_0$ is the zeroth order Bessel function of the first kind.
From above derivation, one sees that the p.d.f. depends on the symmetry group of the theory.

In the chiral limit, Eq. (\ref{fourier}) becomes
\begin{eqnarray}
{\tilde P}(q) \bigg \vert_{m=0} = 
\bigg \langle 
\prod_{j=1}^{N_cV/2} \left( 1- {\left(q 
\over N_cV \lambda_j \right)^2} \right)
\bigg \rangle .
\label{fourier1}
\end{eqnarray}
By computing the second derivative of Eq.(\ref{fourier1}) with respect to $q$,  and
comparing it to the second derivative of $J_0(qc_0)$,
we have the following sum rule:
\begin{eqnarray}
\lim_{m \to 0}  \lim_{V \to \infty} \langle {\bar \psi} \psi \rangle ={c_0} 
= \lim_{V \to \infty} 
\sqrt {{4\over N_c^2V^2} \bigg\langle \sum_{j=1}^{NcV/2} {1\over \lambda_j^2} \bigg\rangle} ,
\label{ps3}
\end{eqnarray}
which agrees with the result from chiral perturbation theory\cite{LS} and chiral
random matrix theory\cite{Ver}.
The advantage is that
no $\lambda$ extrapolation is necessary. 
The disadvantage is that all the eigenvalues of the massless Dirac operator 
have to be calculated as in Eq. (\ref{ps1}). When $V$ is
large, the computational task is huge and not so
feasible.
However, very large lattice is required to get stable and consistent  results.

\subsection{Third relation}

One can perform similar analysis for the p.d.f. of the chiral condensate for a single configuration,
defined in Eq. (\ref{D1}).
The Fourier transformed p.d.f. for this gauge configuration Eq. (\ref{fourier00}) is then
\begin{eqnarray}
{\tilde P}_U(q) &=& \int^{\infty}_{-\infty}
dc ~ \exp \left(-iqc \right) P_U(c)
\nonumber \\
&=&  
\prod_{j=1}^{N_cV/2} \left( 1- {q^2+2imqN_c  V 
\over \left(N_cV\right)^2 \left(\lambda_j^2(U)+m^2\right) } \right) .
\label{D2}
\end{eqnarray}
In the chiral limit, it becomes
\begin{eqnarray}
{\tilde P}_U(q) \bigg \vert_{m=0} = 
\prod_{j=1}^{N_cV/2} \left( 1- \left( {q 
\over N_cV \lambda_j(U) } \right)^2 \right) .
\label{D3}
\end{eqnarray}
This equation is equal to zero at 
\begin{eqnarray}
q=N_cV \lambda_j(U), ~~~j=1, ...,N_cV/2 .
\label{D4}
\end{eqnarray}

Performing similar procedures when deriving Eq. (\ref{C2}), 
one has for ${\tilde P}_U(q) \bigg \vert_{m=0}$
\begin{eqnarray}
{\tilde P}_U(q) \bigg \vert_{m=0}=J_0 \left(qc_0(U)\right) .
\label{D5}
\end{eqnarray}
This equation is equal to zero at 
\begin{eqnarray}
q = {z(j) \over c_0(U)},  ~~~ j=1, ..., \infty , 
\label{D6}
\end{eqnarray}
where $z(j)$ is the $j-$th zero of $J_0$. 
In Tab. \ref{tab1}, the first 40 zeros of $J_0$ 
are provided.

For $V >>1$, Eq. (\ref{D4}) should agree with Eq. (\ref{D6}) so that
\begin{eqnarray}
c_0 (U) = {z(j) \over N_cV \lambda_j (U)},  ~~~ j=1, ..., \infty . 
\label{D7}
\end{eqnarray}
$c_0(U)$ is the amplitude 
of the chiral condensate for configuration $U$. 
Averaging it over gauge configurations with fermions, we obtain
\begin{eqnarray}
C(j)
= 
\langle c_0 (U)  \rangle =
{z(j) \over N_cV} 
\bigg \langle {1 \over \lambda_j} \bigg \rangle .
\label{ps4}
\end{eqnarray}
Neither $m$ nor $\lambda$ extrapolation is necessary.
In the chiral-symmetry breaking phase, 
a plateau for $C(j) = const.$ will develop, from which the chiral condensate
in the chiral limit can be extracted.

The relation between eigenmodes 
and chiral-symmetry breaking is clear:
if chiral symmetry
is spontaneously broken, i.e., 
$c_0(U) \not\equiv 0$, according to Eq. (\ref{D7}) and Eq. (\ref{ps4}),  $\lambda_j$
should scale as $z(j)/V$. 
In the infinite volume limit $V\to \infty$,  the eigenvalues relevant for
chiral-symmetry breaking are those going to zero as
$1/V$, which is consistent with Banks-Casher.

The advantage of Eq. (\ref{ps4}) is that to extract the value of $C(j)$ from a plateau,
only a few smallest eigenvalues are needed 
for this calculation. Of course,
finite size analysis 
$\lim_{m \to 0} \lim_{V \to \infty} <{\bar \psi} \psi> = \lim_{V \to \infty} 
C(j)$
is remained to be done, as in
all approaches.

\section{RESULTS IN QCD WITH KS QUARKS}
\label{sec_Data}

The most interesting application of this method is QCD.
Here we would like to present the first data for SU(3) 
lattice gauge theory with KS fermions, which has the action $S=S_g+S_f$:
\begin{eqnarray}
S_g &=& - {\beta \over N_c} \sum_{p} {\rm Re} ~ {\rm Tr} (U_p) ,
\nonumber \\
S_f &=& \sum_{x,y} {\bar \psi}(x) \Delta_{x,y} \psi (y) , 
\nonumber \\
U_p &=& U_{\mu}(x) U_{\nu}(x+\mu) U_{\mu}^{\dagger} (x+\nu) U_{\nu}^{\dagger} (x) , 
\nonumber \\
\Delta_{x,y}  &=& m \delta_{x,y}
+
\sum_{\mu=1}^4 {1 \over 2} \eta_{\mu}(x)  \left[ U_{\mu}(x) \delta_{x,y-{\hat \mu}} 
-U_{\mu}^{\dagger} (x-{\hat \mu}) \delta_{x,y+{\hat \mu}} \right] ,
\nonumber \\
\eta_{\mu}(x) &=& (-1)^{x_1+x_2+...+x_{\mu-1}} ,
\end{eqnarray}
where $\beta=2 N_c/g^2$.
In the chiral limit $m=0$, 
there exists a U(1) subgroup of the continuous chiral symmetry, i.e.,
$S_f$ is invariant under the following transformation
\begin{eqnarray}
\psi (x) \to \exp\left[ i \alpha (-1)^{x_1+x_2+x_3+x_4}\right] \psi (x) , ~~~
{\bar \psi}(x) \to  {\bar \psi}(x) 
\exp\left[ i \alpha (-1)^{x_1+x_2+x_3+x_4} \right] .
\end{eqnarray}

All simulations are done on the $V=10^4$ lattice
in the quenched SU(3) case. The pure SU(3) gauge fields are updated using the
Cabibbo-Marinari quasi-heat bath algorithm, followed by some over-relaxation sweeps. 100-400 independent configurations
are used for the measurements. 

Figure \ref{fig1} shows the data for $\langle {\bar \psi} \psi \rangle$ at
a stronger coupling $\beta=2.5469$ using Eq. (\ref{ps1}). 
A linear function is also used to extrapolate the data of $\langle {\bar \psi} \psi \rangle$ at nonzero fermion mass 
$m \in [0.005,0.1]$ to the chiral limit.
Figure \ref{fig2} shows the result of $C(j)$ for the same 
$\beta$ using Eq. (\ref{ps4}); there is a nice
plateau for $j \in [15,40]$. The linear extrapolation result from Eq. (\ref{ps1}) is
consistent with the mean value of Eq. (\ref{ps4}) in the plateau.

The results at a weaker coupling $\beta=5.6263$ are shown 
in Figs. \ref{fig3} and \ref{fig4}. 
One sees there is a little change in the slope when using Eq. (\ref{ps1})
to calculate $\langle {\bar \psi} \psi \rangle$ and the extrapolated value at $m=0$.
In comparison, there is a wide plateau for $C(j)$ using Eq. (\ref{ps4}).
But both approaches still give consistent results.

In conclusion, 
we have shown how the p.d.f. method
for obtaining the chiral condensate 
in the exact chiral limit works in lattice QCD with quenched KS quarks.
There are several advantages in using relation Eq. (\ref{ps4}): 
only calculations of a small set of eigenvalues of the massless Dirac operator
are necessary; there is no need for $m$ or $\lambda$ extrapolation. 
This might be an alternative efficient method for investigating the spontaneous
chiral-symmetry breaking in lattice QCD. 
It would also be very interesting to see application of this approach to other fermion formulations, e.g.,
overlap fermions or domain wall fermions.

\vskip 1cm
\noindent
{\bf Acknowledgments}


I am grateful to V. Azcoiti and V. Laliena for useful discussions and collaboration at the beginning of the work.
This work is supported by the 
Key Project of National Science Foundation (10235040), 
National and Guangdong Ministries of Education, and
Foundation of the Zhongshan University 
Advanced Research Center.

\textheight 9.0in

\begin{table}
\begin{center}
  \begin{tabular}{|c|c|}\hline
$j$ & $z(j)$  \\ \hline
1 & 2.4048 \\
2 & 5.5200 \\
3 & 8.6540 \\
4 & 11.7920 \\
5 & 14.9310 \\
6 & 18.0710 \\
7 & 21.2120 \\
8 & 24.3530 \\
9 & 27.4940 \\
10 & 30.6346 \\
11 & 33.7758 \\
12 & 36.9171 \\
13 & 40.0584 \\
14 & 43.1998 \\
15 & 46.3412 \\
16 & 49.4826 \\
17 & 52.6241 \\
18 & 55.7655 \\
19 & 58.9070 \\
20 & 62.0485 \\
21 & 65.1900 \\
22 & 68.3315 \\
23 & 71.4730 \\
24 & 74.6145 \\
25 & 77.7562 \\
26 & 80.8976 \\
27 & 84.0391 \\
28 & 87.1806 \\
29 & 90.3222 \\
30 & 93.4637 \\
31 & 96.6053 \\
32 & 99.7468 \\
33 & 102.8884 \\
34 & 106.0299 \\
35 & 109.1715 \\
36 & 112.3131 \\
37 & 115.4546 \\
38 & 118.5962 \\
39 & 121.7377 \\
40 & 124.8793 \\
\hline
\end{tabular}
\end{center}
\caption{\label{tab1} First 40 zeros of $J_0$.}
\end{table}

\input figure_df9010.tex

\end{document}